# Bunched proton acceleration from a laser-irradiated cone target

Xing-Long Zhu[1, 2, 3,*, ‡], Wei-Yuan Liu[2, 3, ‡], Min Chen[2, 3], Su-Ming Weng[2, 3], Paul McKenna[4,5], Zheng-Ming Sheng[1,2,3,4,5*], and Jie Zhang[1, 2, 3]

[1] Tsung-Dao Lee Institute, Shanghai Jiao Tong University, Shanghai 200240, China
[2] Key Laboratory for Laser Plasmas (MOE), School of Physics and Astronomy, Shanghai Jiao Tong University, Shanghai 200240, China
[3] Collaborative Innovation Center of IFSA, Shanghai Jiao Tong University, Shanghai 200240, China
[4] SUPA, Department of Physics, University of Strathclyde, Glasgow G4 0NG, United Kingdom
[5] Cockcroft Institute, Sci-Tech Daresbury, Cheshire WA4 4AD, United Kingdom
‡ These authors contributed equally to this work.
* Email: xinglong.zhu@sjtu.edu.cn (X.L.Z.); zmsheng@sjtu.edu.cn (Z.M.S.)

## Abstract

**Laser-driven ion acceleration is an attractive technique for compact high-energy ion sources. Currently, among various physical and technical issues to be solved, the boost of ion energy and the reduction of energy spread represent the key challenges with this technique. Here we present a scheme to tackle these challenges by using a hundred-terawatt-class laser pulse irradiating a cone target. Three-dimensional particle-in-cell simulations show that a large number of electrons are dragged out of the cone walls and accelerated to hundreds of MeV by the laser fields inside the cone. When these energetic dense electron beams pass through the cone target tip into vacuum, a very high bunching acceleration field, up to tens of TV/m, quickly forms. Protons are accelerated and simultaneously bunched by this field, resulting in quasi-monoenergetic proton beams with hundred MeV energy and low energy spread of ~2%. Results exploring the scaling of the proton beam energy with laser and target parameters are presented, indicating that the scheme is robust. This opens a new route for compact high-energy proton sources from fundamental research to biomedical applications.**

High-energy ion sources generated by intense laser-plasma interactions have attracted significant interest in the past two decades [1-4] due to their potential applications, including in fast ignition fusion [5-7], radiography for laser-matter interaction and ultrafast dynamics [8-10], tabletop ion accelerators [11], radioisotope production for nuclear physics [12], and generation of high-energy-density matter [13]. They can also potentially be applied in cancer therapy and medical diagnosis [14-17]. Many of these applications require or benefit from quasi-monoenergetic ion beams with low energy spread, down to 1% level for tumor therapy, for example, in addition to the energy as high as tens-to-hundreds of MeV per nucleon. Significant advancements have been achieved in boosting the ion energy generated in laser-ion acceleration [18-20], with maximum proton energy over 94 MeV recently demonstrated using a few-hundred-Joule, picosecond laser [21]. Such high energy laser facilities normally operate with a low-repetition rate and are housed at large national laboratories. Remarkable development in high-power femtosecond laser and target technologies offers possibilities for compact energetic ion sources. Several advances have been made in recent experiments using petawatt (PW) lasers with tens-of-J energy [22-24]. However, the ion energy spectra are still exponential-like rather than monoenergetic or quasi-monoenergetic, restricting their use for potential applications.

To achieve compact monoenergetic ion sources, developing a large bunching acceleration field is crucial. The target normal sheath acceleration (TNSA) is one of the most widely-investigated laser-ion acceleration mechanisms [25-30]. However, TNSA-produced ion beams are characterized by exponential energy spectra. Moreover, the maximum energy of ions produced scales with the laser pulse energy [3, 4, 31], which is unfavorable for the use of low-energy femtosecond laser drivers. Another widely-studied mechanism is radiation pressure acceleration (RPA) [32-38], where the pressure exerted by the intense laser light drives plasma electrons inwards, giving rise to a charge-separation electric field and thus ion acceleration. Although this is a promising scheme at ultrahigh intensities, the maximum ion energy is limited in the sub-PW regime since strict requirements of ultrahigh-contrast pulses and laser polarization. Moreover, the energy spread is affected by transverse instabilities [39, 40] and the finite laser spot size [41], leading to rapid deformation of the targets and adversely affecting the ion bunching acceleration field. To improve the maximum ion energy and/or energy spread, a variety of approaches have been introduced, including the use of multi-pulse or multi-stage acceleration schemes [42-45], specially designed targets [46-52], other novel laser and target parameters [53-56], and post-acceleration [57]. Despite the progress made, the ion beam energy spread typically remains above 10%. Very recently, a scheme has been proposed to produce 100 MeV proton beams with a few percent energy spread by a PW intense laser irradiating a microtape target [58]. Efforts continue to extend laser-ion acceleration to higher energies, smaller energy spreads and higher repetition rates at routinely available laser powers, which is strongly needed for the development of future compact ion accelerators and practical applications, enabling stable high-energy ion sources to a broad community.

To address this quest, we present a scheme to generate quasi-monoenergetic proton beams via the interaction of a readily available 100-terawatt (TW) class laser with a cone target. In this scheme, longitudinal bunching and transverse focusing develop during proton acceleration due to the unique high field structure induced by excessive energetic electrons accelerated from the cone, leading to the generation of hundred-MeV proton beams with low energy spreads of a few percent. The advantage of this scheme over approaches previously explored is that it can simultaneously maximize the proton peak energy and minimize its energy spread, and is achievable with relatively low peak power lasers. The findings open a new way for the development of compact high-energy ion sources.

Figure 1(a) illustrates the scheme for generating quasi-monoenergetic proton beams. As the laser propagates into the cone, a large number of electrons are periodically extracted from the cone walls in the transverse direction and are accelerated by the laser fields in the longitudinal direction, where the cone guides them forward, as illustrated in Fig. 1(b). Meanwhile, light intensity is enhanced due to laser focusing within the cone, intensifying high-energy electron generation. When such dense energetic electron beams pass through the target tip into vacuum, a longitudinal field is established at the rear of the target, which leads to simultaneous proton acceleration and bunching owing to excessive huge electron charge (above 100nC), as shown in Fig. 1(c). This results in the generation of a monoenergetic proton beam with near 100-MeV peak energy, approximately 2% energy spread, and proton number exceeding $10^9$, when driven by a relativistic femtosecond laser with only a few Joules of energy.

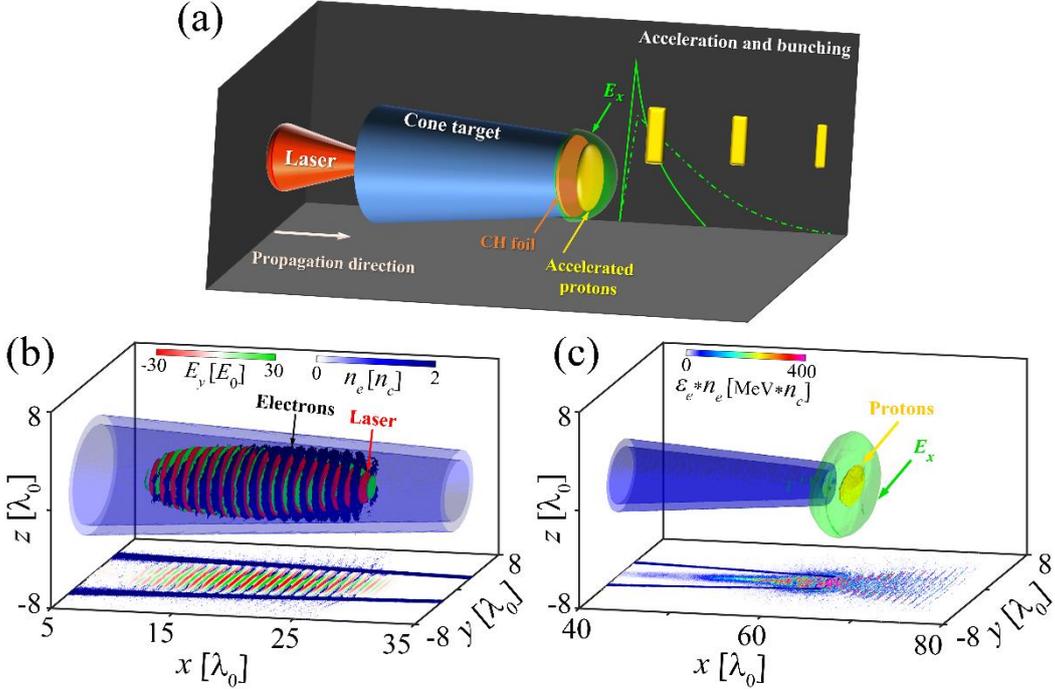

**Figure 1.** (a) Schematic of laser-irradiated cone acceleration for quasi-monoenergetic proton beams. A 100-TW-class femtosecond laser is incident on the left aperture of a cone target, where a thin hydrocarbon (CH) foil is attached at the cone apex. With the laser pulse passing through the cone, its intensity is greatly enhanced owing to laser focusing effects, so that large numbers of electrons are dragged out of the cone walls and accelerated by the enhanced laser fields, as seen in (b). Later, these dense bunches of energetic electrons traverse the end wall of the cone target into vacuum and set up a strong longitudinal electrostatic field, as seen in (c). This field is ideal to simultaneously accelerate and bunch protons, producing a quasi-monoenergetic proton beam of over hundred MeV. The green solid and dashed lines in (a) illustrate schematically the evolution of accelerating sheath fields, which can lead to proton bunching as illustrated by the bars in yellow.

To demonstrate the proposed scheme, we carry out three-dimensional (3D) particle-in-cell (PIC) simulations with the electromagnetic relativistic code EPOCH [59]. The size of the simulation box is $100\lambda_0(x) \times 16\lambda_0(y) \times 16\lambda_0(z)$ with $3500 \times 400 \times 400$ grid cells, where 27 macro-particles in each cell are used for both ions and electrons. The incident linearly-polarized laser pulse has a spatial Gaussian distribution and a $\sin^2$ temporal profile, with spot radius $\sigma_0 = 3\lambda_0$, wavelength $\lambda_0 = cT_0 = 0.8\mu m$, and pulse duration of $\tau_0 = 29fs$ full width at half maximum (FWHM). The normalized laser amplitude is $a_0 = \frac{eE_y}{m_e c \omega_0} = 36$, with about 7J pulse energy and $P_0 = 250TW$ power, which can be delivered by commercially available hundreds-of-TW laser systems, where $\omega_0$ is the laser oscillation frequency, $E_y$ is the electric field, $e$ is the elementary charge, $m_e$ is the electron mass, and $c$ is the speed of light in vacuum. The solid cone target (assuming aluminum) has a $60\lambda_0$ axially-longitudinal length, with a left aperture radius of $R_0 = 4\lambda_0$ and a right aperture radius of $r_0 = 1\lambda_0$, where a CH layer (i.e. a thin plastic foil) with thickness $d_0 = 0.2\lambda_0$ is placed at its apex to provide a source of protons. The charge states of ions are initialized as $Al^{13+}$ for the cone target, and a uniform mixture of $H^+$ and $C^{6+}$ for the CH foil, both targets with $n_0 = 90n_c$ electron densities, where $n_c = \frac{m_e \omega_0^2}{4\pi e^2}$ is the critical density. For the purpose of implementing the scheme in experiments, the target can also be made of other materials, and it may have a larger inlet size, as long as the laser pulse is focused into the cone. It should be mentioned that our scheme is different from the previous laser-cone interactions for ion energy enhancement [60-66], where the ions are still accelerated via TNSA and have exponential energy spectra, and also different from other studies using the conical targets such as laser fusion ignition [67-69] and QED effects [70-72].

The key to attaining monoenergetic ion beams is to create a negative-gradient acceleration field, which sustains for a long time and is large enough so that proton bunching develops. Although the RPA process can form such a field, it only lasts for a short period of time, as discussed earlier. To fulfill these conditions, we employ a cone target with a thin CH foil attached at its apex to make two efficient stages for high-energy dense electron beams and monoenergetic proton beams, respectively.

In the first stage, the cone plays an important role in efficient electron acceleration and laser propagation, where laser pulse is guided over a distance much longer than the Rayleigh length. As the laser propagates through the cone, its electric field ($E_y$) is sufficiently strong to drag large numbers of electrons directly out of the wall inner surface [Fig. 2(a)]. Over each laser period, a positive $E_y$ drags electrons from the upper wall (y>0) in the first half of cycle, while a negative $E_y$ extracts electrons from the lower wall in the next half of cycle, where the pulled electrons attain relativistic velocities within half of a laser period. These electrons are accelerated in the forward direction by the $-e\mathbf{V} \times \mathbf{B}_z$ force and are further accelerated by direct laser acceleration [73-75].

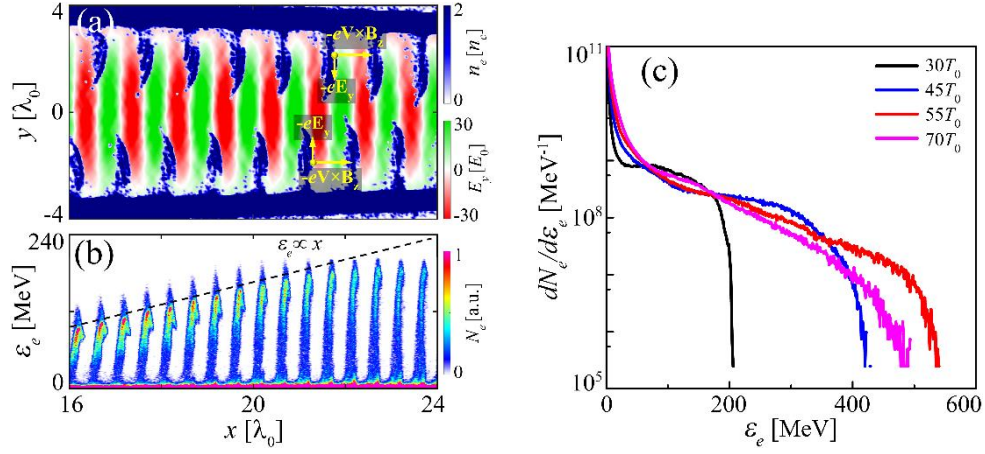

**Figure 2.** (a) Distributions of the laser field ($E_y$) and electron density ($n_e$) at $t = 30T_0$, where electrons are dragged from the cone wall and accelerated in the forward direction. (b) Distribution of the electron energy space ($x, \varepsilon_e$) at $t = 30T_0$, where the black dashed line shows the scaling of maximum electron energy. (c) The energy spectrum of electrons at stated times.

The energy gain of the electrons during the acceleration can be estimated by $\varepsilon_e \approx e\bar{E}_x l_d$, where $\bar{E}_x$ is the averge longitudinal electric field and $l_d$ is the effective accelerating distance. One may further derive the longitudinal electric field amplitude $E_x = -\frac{2y}{k\sigma_0^2} a_0 E_0 \exp(-\frac{r^2}{\sigma_0^2})$ in the first-order approximation and obtain its average value $\bar{E}_x \approx \frac{a_0 E_0}{k\sigma_0}$, where $E_0 = \frac{m_e c \omega_0}{e}$. Since strong focusing of the laser pulse occurs within the cone, both its intensity and $E_x$ are greatly enhanced. The maximum electron energy can be calculated as

$$\varepsilon_{e,m} \approx \bar{a}_{foc} l_d m_e c^2 / \sigma_0, \tag{1}$$

where $\bar{a}_{foc}$ is the average focusing laser amplitude. This suggests that the electron energy increases with $l_d$, which is validated by 3D PIC simulation [see Fig. 2(b)]. Taking $l_d \sim 60\lambda_0$ and $\bar{a}_{foc} \sim 60$ in our configuration, one can predict the maximum electron energy to be about 600MeV, which is consistent with the results presented in Fig. 2(c). Later, electrons lose energy due to their excitation of strong charge-separation fields, but their effective temperature is still over 100MeV, which is much higher than the ponderomotive scaling [76] $\left(\sqrt{1+a_0^2}-1\right)m_e c^2 \approx 18$MeV and other target configurations with typically a few to tens of MeV.

In the second stage, when the overdense energetic electron beams arrive at the apex of the cone, a strong charge-separation field is formed. One may estimate the longitudinal sheath field as [77, 78]

$$E_s \approx \sqrt{\frac{2}{ec}\frac{T_h}{eL_s}}, \tag{2}$$

where $L_s \sim \sqrt{T_h/4\pi e^2 n_h}$ and $e_C \approx 2.71828$, $T_h$ and $n_h$ are respectively the electron temperature and density. By substituting $T_h \sim 120 \text{MeV}$ and $n_h \sim 2n_c$ into Eq.(2), the field peak is estimated to be about $7 \times 10^{13} \text{V/m}$, which agrees with the simulation result $6 \times 10^{13} \text{V/m}$. It is about six orders of magnitude higher than that achieved in conventional accelerators, enabling efficient ion acceleration. Protons are pulled from the CH target and accelerated by such an extremely intense field. Meanwhile, they are surrounded by a large electron cloud, where the proton number (a few nC) is much lower than that of energetic electrons (about 130 nC with energy >5MeV). This causes $\partial E_x/\partial x = 4\pi e(n_p - n_e) < 0$. Therefore, a transverse focusing electrostatic field $E_z$ [Fig. 3(a)] and a negative-gradient longitudinal bunching field $E_x$ [Fig. 3(b)] are established synchronously. Such bunching and focusing fields are very helpful to attain monoenergetic proton beams. The field $E_x$ longitudinally accelerates protons and compresses their phase space [see Figs. 3(c) and 3(d)], while $E_z$ transversely focuses protons towards the center, such that the proton energy spread is immensely reduced to about 2% (FWHM). Finally, it gives a highly monoenergetic proton beam with near 100MeV peak energy and approximately $2 \times 10^9$ number within the peak, as seen in Fig. 3(e).

As a comparison, we consider the case of the same laser pulse directly-irradiating a thin CH foil without the cone target [see the inset in Fig. 3(e)], i.e. a RPA-like case, while keeping all other parameters fixed. In this case, due to the lack of effective electron acceleration, a sufficient number of energetic electrons and a long-lasting huge sheath field, the bunching acceleration field disappears earlier and its strength much lower in the laser-foil case. The consequence is that the proton energy spectrum is quite broad and its peak is only about 53MeV. Therefore, our scheme is different from the RPA one and is more suitable for achieving monoenergetic ion beams. It should be mentioned that the acceleration field can last well over 100fs and has a length of tens of microns at the cone target rear, although its strength is reduced by about an order of magnitude at $t = 100T_0$. It later develops a dual-peaked structure due to an electrostatic field (the lower peak) excited by the bunched proton beam, which is beneficial for eliminating the overly high phase velocity in the beam tail.

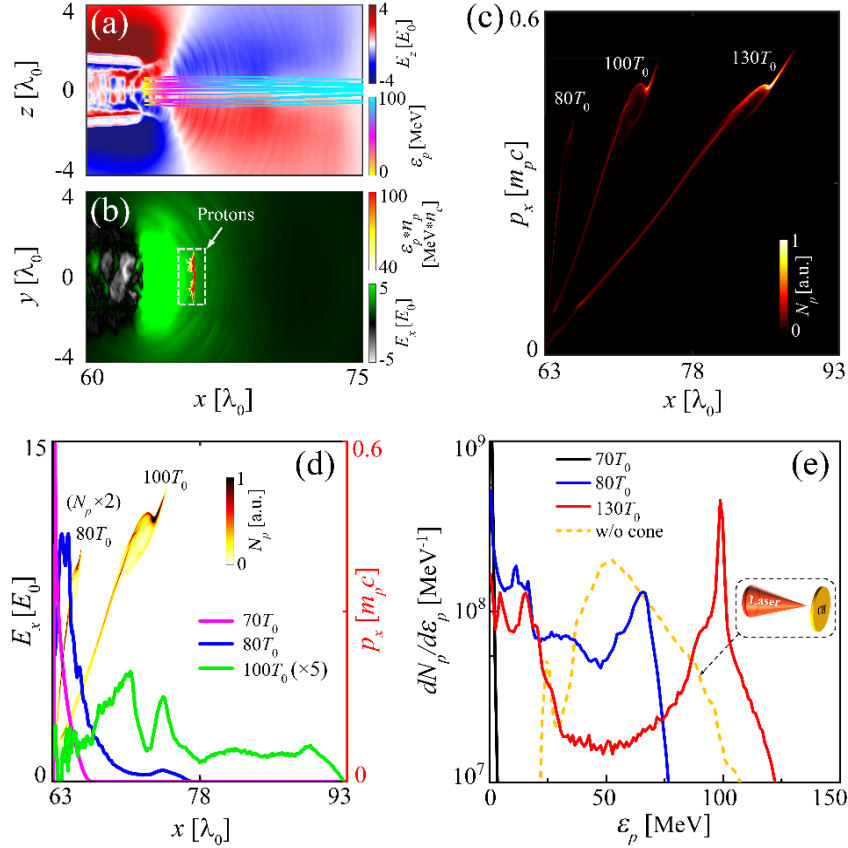

**Figure 3.** Distributions of (a) the transverse field ($E_z$), and (b) the longitudinal field ($E_x$) and proton energy density at $t = 80T_0$. The colored lines in (a) represent the trajectories of some typical protons. (c) Evolution of the proton phase space. (d) Evolution of the on-axis spatial distribution of $E_x$. For reference, the maps also depict the proton phase space at $t = 80T_0$ and $100T_0$. (e) The energy spectrum of protons at stated times, where the yellow dashed line indicates the result from laser-driven a thin CH foil.

In order for large numbers of protons to be steadily accelerated and bunched, the electrostatic field induced by the protons should be lower than the sheath field $E_s$, which can be described as $4\pi e n_p d \leq E_s$ in a simple one-dimensional model. By substituting $n_p d \approx n_{p0} d_0 = n_0 d_0 / 7$, $n_c = \frac{m_e \omega_0^2}{4\pi e^2}$ and $E_0 = \frac{m_e c \omega_0}{e}$ into this model, the optimal bunching condition can be expressed as

$$E_s \gtrsim 2\pi E_0 n_0 d_0 / 7 n_c \lambda_0, \tag{3}$$

where $n_0$ and $d_0$ are respectively the initial electron density and thickness of the CH layer, and the numerical value 7 is attributed to the CH target. Its form is similar to that of RPA, but the underlying physics is different. In our scheme, the acceleration and bunching processes are dominated by the long-pulse electron beam-excited sheath field rather than the laser light pressure, and the limitations in RPA such as transverse instability and finite laser spot size effects are avoided. Also this is different from the previous laser-cone interaction cases [60-66] where the ions are mostly accelerated via the normal TNSA and have typically exponential-like energy spectra due to the lack of bunching and focusing fields. The effect of the product of the foil thickness and density (where $d_0 = 0.2\lambda_0$ is fixed) on the proton energy and spread is shown in Fig. 4(a). The resulting peak energy (almost around 100MeV) mainly depends on $E_s$ instead of $n_0 d_0$ in RPA, while energy spread can be optimized as $E_{s0} \sim 2\pi E_0 n_0 d_0 / 7 n_c \lambda_0$. Accordingly, the optimal target parameter should be $n_0 d_0 \sim 16.7 n_c \lambda_0$ for $E_{s0} \approx 15 E_0$, which agrees well with the simulation results. Furthermore, we can infer that for a fixed target $E_s \gtrsim E_{s0}/2$ is needed in order for the proton energy spread to be much less than 10%, which corresponds roughly to the laser power $\gtrsim P_0/4$ since $E_s \propto \sqrt{I} \propto \sqrt{P}$. Taking the parameters ($P_0 = 250$TW) in Fig. 3 for example, the minimum laser power required is around 62TW, which is testified by our simulations depicted in Fig. 4(c).

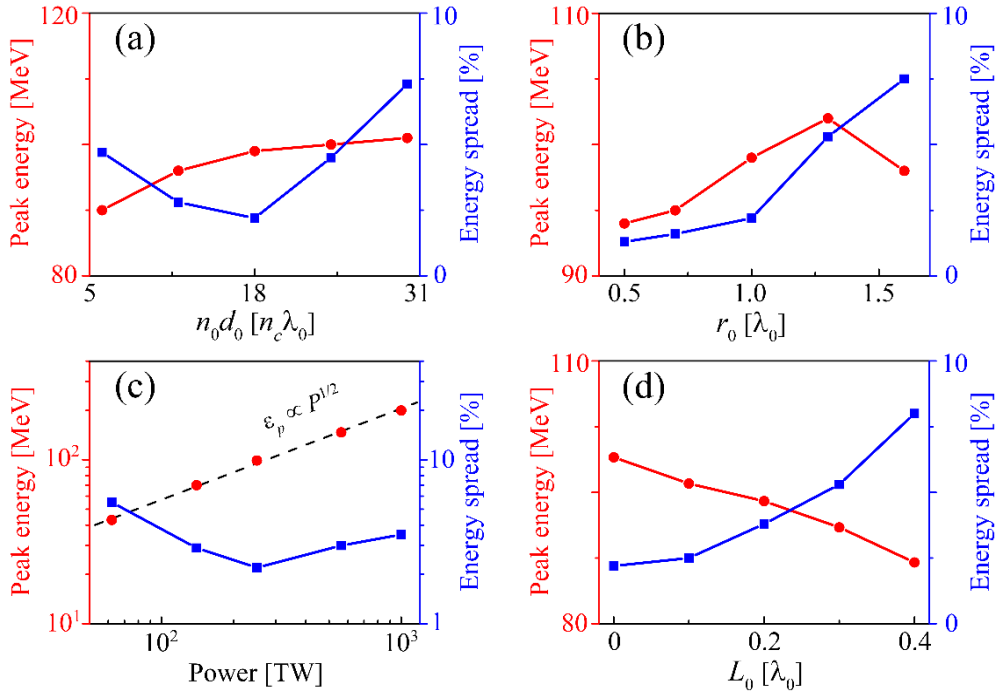

**Figure 4.** Effects of (a) the product of foil density and thickness ($n_0 d_0$), (b) cone apex aperture radius ($r_0$), (c) laser power ($P$), and (d) preplasma scale length ($L_0$) on the proton peak energy and energy spread. The black dashed line in (c) presents the scaling of proton peak energy with laser power.

One can further manipulate the proton peak energy and spread by varying the laser and cone parameters. We first discuss the effect of the apex aperture size of the cone target. Figure 4(b) presents the results of changing $r_0$ from 0.5 to $1.6\lambda_0$, where the other parameters are fixed. It is shown that an appropriately small apex aperture is conducive to diminishing the proton energy spread. For example, for a cone target with $r_0 = 0.5\lambda_0$, proton beams with a very low energy spread of about 1% and 94 MeV peak energy can be obtained. This is because a large number of electrons are guided and concentrated at the cone apex, where those around the propagation axis increase as the apex size decreases, which benefits the bunching and focusing fields and hence can improve the proton energy spread. Nevertheless, the aperture size should not be too small; otherwise, the laser pulse will deplete prematurely and high-energy electrons will reduce, causing a decrease in the longitudinal accelerating field and thus lower proton energy. Overall, an optimal aperture radius should be around $1\lambda_0$, which can achieve 100 MeV peak energy while maintaining about 2% level energy spread.

The effect of the laser power on the ion beam generation is illustrated in Fig. 4(c), where the power is altered in the range of 62TW to 1PW, while the other parameters remain almost the same, except the cone target density to avoid laser-induced relativistic transparency at high intensities. The created proton energy can be estimated by $\varepsilon_p \approx eE_sL_s \sim T_h$ according to Eq. (2). Here the electron temperature can be scaled as $T_h \propto a_0$ following the energy scaling $\varepsilon_e \propto a_0$. When the incident laser geometry is fixed, the maximum intensity is determined by its power as $P = \pi\sigma_0^2 I_0/2$, where $I_0 = \frac{\pi}{2}\frac{m_e^2 c^5}{e^2 \lambda_0^2}a_0^2$. Thus, the scaling of the proton peak energy can be roughly expressed as

$$\varepsilon_p \sim \alpha\sqrt{P/P_{100}}, \qquad (4)$$

where $\alpha \approx 63$MeV is the coefficient and $P_{100} = 100$TW, which is verified by the simulation results shown in Fig. 4(c). Although it is similar to TNSA scaling, the magnitude of proton energy is greatly higher for a given laser power, which is attributed to a super-strong bunching acceleration field excited by hundreds-of-MeV dense electron beams. For example, when using a 1PW laser, the peak energy of protons can reach 200 MeV. More importantly, the energy spreads of proton beams obtained are quite low just a few percent, even in the 62 TW laser case, it is still less than 6%. Therefore, this is a robust and efficient scheme to produce quasi-monoenergetic ion beams with energies ranging from tens-to-hundreds of MeV, which would help to enable the development of compact ion sources with high-repetition-rate terawatt-class lasers, enhancing their availability and stability.

We also investigate the robustness of this scheme in terms of the preplasmas in Fig. 4(d). The preplasmas at the cone walls and the front of the CH-foil have a linear density ramp with the scale length $L_0$ ranging from 0 to $0.4\lambda_0$, and have $2\lambda_0$ linear density ramp at the front of the cone walls, where the other parameters are fixed. The results show that our scheme can allow a larger preplasma than RPA, which will significantly reduce the laser contrast requirement. Also there is no requirement for the use of circularly-polarized lasers in our scheme. All these make it robust for quasi-monoenergetic proton acceleration.

In conclusion, we propose and numerically demonstrate an efficient scheme to generate hundred MeV proton beams with low energy spreads, at a few percent level, from a cone target irradiated by 100 TW-class intense femtosecond lasers. These findings show that strong longitudinal acceleration fields and transverse focusing fields are generated when high currents of electrons pass from the cone target tip into vacuum. Simultaneous proton acceleration and bunching are found at the front of the huge sheath field, enabling quasi-monoenergetic proton beam generation. This scheme paves the way towards next-generation compact ion sources with tens-to-hundreds of MeV energies, offering new opportunities for various applications from fundamental research to medicine.

**Acknowledgements**
We acknowledge the support from the Strategic Priority Research Program of Chinese Academy of Sciences (Grant No. XDA25050100), the National Postdoctoral Program for Innovative Talents (Grant No. BX20220206), the National Natural Science Foundation of China (Grant Nos. 11991074, 11975154, and 12135009), Science Challenge Project (No. TZ2018005), and the core grant of the Cockcroft Institute from the UK STFC (Grant No. ST/V001612/1). The development of EPOCH code was supported in part by the UK EPSRC (Grant No. EP/G056803/1). Numerical simulations were carried out on the PI2 supercomputer at Shanghai Jiao Tong University.

**References**
[1] A. Macchi, M. Borghesi, M. Passoni, Ion acceleration by superintense laser-plasma interaction, *Rev. Mod. Phys.* **85**, 751 (2013).
[2] H. Daido, M. Nishiuchi, A. S. Pirozhkov, Review of laser-driven ion sources and their applications, *Rep. Prog. Phys.* **75**, 056401 (2012).
[3] J. Fuchs, P. Antici, E. d'Humières, E. Lefebvre, M. Borghesi, E. Brambrink, C. A. Cecchetti, M. Kaluza, V. Malka, M. Manclossi *et al.*, Laser-driven proton scaling laws and new paths towards energy increase, *Nat.*

*Phys.* **2**, 48 (2006).

[4] L. Robson, P. T. Simpson, R. J. Clarke, K. W. D. Ledingham, F. Lindau, O. Lundh, T. McCanny, P. Mora, D. Neely, C. G. Wahlström *et al.*, Scaling of proton acceleration driven by petawatt-laser–plasma interactions, *Nat. Phys.* **3**, 58 (2007).

[5] M. Roth, T. E. Cowan, M. H. Key, S. P. Hatchett, C. Brown, W. Fountain, J. Johnson, D. M. Pennington, R. A. Snavely, S. C. Wilks *et al.*, Fast Ignition by Intense Laser-Accelerated Proton Beams, *Phys. Rev. Lett.* **86**, 436 (2001).

[6] S. Atzeni, M. Temporal, J. J. Honrubia, A first analysis of fast ignition of precompressed ICF fuel by laser-accelerated protons, *Nucl. Fusion* **42**, L1 (2002).

[7] B. M. Hegelich, D. Jung, B. J. Albright, J. C. Fernandez, D. C. Gautier, C. Huang, T. J. Kwan, S. Letzring, S. Palaniyappan, R. C. Shah *et al.*, Experimental demonstration of particle energy, conversion efficiency and spectral shape required for ion-based fast ignition, *Nucl. Fusion* **51**, 083011 (2011).

[8] C. K. Li, F. H. Séguin, J. A. Frenje, J. R. Rygg, R. D. Petrasso, R. P. J. Town, P. A. Amendt, S. P. Hatchett, O. L. Landen, A. J. Mackinnon *et al.*, Measuring E and B Fields in Laser-Produced Plasmas with Monoenergetic Proton Radiography, *Phys. Rev. Lett.* **97**, 135003 (2006).

[9] G. Sarri, C. A. Cecchetti, L. Romagnani, C. M. Brown, D. J. Hoarty, S. James, J. Morton, M. E. Dieckmann, R. Jung, O. Willi *et al.*, The application of laser-driven proton beams to the radiography of intense laser–hohlraum interactions, *New J. Phys.* **12**, 045006 (2010).

[10] B. Dromey, M. Coughlan, L. Senje, M. Taylor, S. Kuschel, B. Villagomez-Bernabe, R. Stefanuik, G. Nersisyan, L. Stella, J. Kohanoff *et al.*, Picosecond metrology of laser-driven proton bursts, *Nat. Commun.* **7**, 10642 (2016).

[11] T. E. Cowan, J. Fuchs, H. Ruhl, A. Kemp, P. Audebert, M. Roth, R. Stephens, I. Barton, A. Blazevic, E. Brambrink *et al.*, Ultralow Emittance, Multi-MeV Proton Beams from a Laser Virtual-Cathode Plasma Accelerator, *Phys. Rev. Lett.* **92**, 204801 (2004).

[12] K. W. D. Ledingham, P. McKenna, R. P. Singhal, Applications for Nuclear Phenomena Generated by Ultra-Intense Lasers, *Science* **300**, 1107 (2003).

[13] P. K. Patel, A. J. Mackinnon, M. H. Key, T. E. Cowan, M. E. Foord, M. Allen, D. F. Price, H. Ruhl, P. T. Springer, R. Stephens, Isochoric Heating of Solid-Density Matter with an Ultrafast Proton Beam, *Phys. Rev. Lett.* **91**, 125004 (2003).

[14] S. V. Bulanov, T. Z. Esirkepov, V. S. Khoroshkov, A. V. Kuznetsov, F. Pegoraro, Oncological hadrontherapy with laser ion accelerators, *Phys. Lett. A* **299**, 240 (2002).

[15] V. Malka, S. Fritzler, E. Lefebvre, E. d'Humières, R. Ferrand, G. Grillon, C. Albaret, S. Meyroneinc, J.-P. Chambaret, A. Antonetti *et al.*, Practicability of protontherapy using compact laser systems, *Med. Phys.* **31**, 1587 (2004).

[16] S. Fritzler, V. Malka, G. Grillon, J. P. Rousseau, F. Burgy, E. Lefebvre, E. d'Humières, P. McKenna, K. W. D. Ledingham, Proton beams generated with high-intensity lasers: Applications to medical isotope production, *Appl. Phys. Lett.* **83**, 3039 (2003).

[17] S. Kawata, T. Izumiyama, T. Nagashima, M. Takano, D. Barada, Q. Kong, Y. J. Gu, P. X. Wang, Y. Y. Ma, W. M. Wang, Laser ion acceleration toward future ion beam cancer therapy - Numerical simulation study, *Laser therapy* **22**, 103 (2013).

[18] R. A. Snavely, M. H. Key, S. P. Hatchett, T. E. Cowan, M. Roth, T. W. Phillips, M. A. Stoyer, E. A. Henry, T. C. Sangster, M. S. Singh, Intense high-energy proton beams from petawatt-laser irradiation of solids, *Phys. Rev. Lett.* **85**, 2945 (2000).

[19] S. A. Gaillard, T. Kluge, K. A. Flippo, M. Bussmann, B. Gall, T. Lockard, M. Geissel, D. T. Offermann, M. Schollmeier, Y. Sentoku *et al.*, Increased laser-accelerated proton energies via direct laser-light-pressure acceleration of electrons in microcone targets, *Phys. Plasmas* **18**, 056710 (2011).

[20] F. Wagner, O. Deppert, C. Brabetz, P. Fiala, A. Kleinschmidt, P. Poth, V. A. Schanz, A. Tebartz, B. Zielbauer, M. Roth *et al.*, Maximum Proton Energy above 85 MeV from the Relativistic Interaction of Laser Pulses with Micrometer Thick CH2 Targets, *Phys. Rev. Lett.* **116**, 205002 (2016).

[21] A. Higginson, R. J. Gray, M. King, R. J. Dance, S. D. R. Williamson, N. M. H. Butler, R. Wilson, R. Capdessus, C. Armstrong, J. S. Green *et al.*, Near-100 MeV protons via a laser-driven transparency-enhanced hybrid acceleration scheme, *Nat. Commun.* **9**, 724 (2018).

[22] P. Wang, Z. Gong, S. G. Lee, Y. Shou, Y. Geng, C. Jeon, I. J. Kim, H. W. Lee, J. W. Yoon, J. H. Sung *et al.*, Super-Heavy Ions Acceleration Driven by Ultrashort Laser Pulses at Ultrahigh Intensity, *Phys. Rev. X* **11**, 021049 (2021).

[23] I. J. Kim, K. H. Pae, I. W. Choi, C.-L. Lee, H. T. Kim, H. Singhal, J. H. Sung, S. K. Lee, H. W. Lee, P. V. Nickles *et al.*, Radiation pressure acceleration of protons to 93 MeV with circularly polarized petawatt laser pulses, *Phys. Plasmas* **23**, 070701 (2016).

[24] T. Ziegler, D. Albach, C. Bernert, S. Bock, F. E. Brack, T. E. Cowan, N. P. Dover, M. Garten, L. Gaus, R. Gebhardt *et al.*, Proton beam quality enhancement by spectral phase control of a PW-class laser system, *Sci. Rep.* **11**, 7338 (2021).

[25] A. Pukhov, Three-Dimensional Simulations of Ion Acceleration from a Foil Irradiated by a Short-Pulse Laser, *Phys. Rev. Lett.* **86**, 3562 (2001).

[26] S. C. Wilks, A. B. Langdon, T. E. Cowan, M. Roth, M. Singh, S. Hatchett, M. H. Key, D. Pennington, A. MacKinnon, R. A. Snavely, Energetic proton generation in ultra-intense laser–solid interactions, *Phys. Plasmas* **8**, 542 (2001).

[27] H. Schwoerer, S. Pfotenhauer, O. Jäckel, K. U. Amthor, B. Liesfeld, W. Ziegler, R. Sauerbrey, K. W. D. Ledingham, T. Esirkepov, Laser-plasma acceleration of quasi-monoenergetic protons from microstructured targets, *Nature* **439**, 445 (2006).

[28] B. M. Hegelich, B. J. Albright, J. Cobble, K. Flippo, S. Letzring, M. Paffett, H. Ruhl, J. Schreiber, R. K. Schulze, J. C. Fernández, Laser acceleration of quasi-monoenergetic MeV ion beams, *Nature* **439**, 441 (2006).

[29] M. Nakatsutsumi, Y. Sentoku, A. Korzhimanov, S. N. Chen, S. Buffechoux, A. Kon, B. Atherton, P. Audebert, M. Geissel, L. Hurd *et al.*, Self-generated surface magnetic fields inhibit laser-driven sheath acceleration of high-energy protons, *Nat. Commun.* **9**, 280 (2018).

[30] N. P. Dover, M. Nishiuchi, H. Sakaki, K. Kondo, M. A. Alkhimova, A. Y. Faenov, M. Hata, N. Iwata, H. Kiriyama, J. K. Koga *et al.*, Effect of Small Focus on Electron Heating and Proton Acceleration in Ultrarelativistic Laser-Solid Interactions, *Phys. Rev. Lett.* **124**, 084802 (2020).

[31] J. Schreiber, F. Bell, F. Grüner, U. Schramm, M. Geissler, M. Schnürer, S. Ter-Avetisyan, B. M. Hegelich, J. Cobble, E. Brambrink *et al.*, Analytical Model for Ion Acceleration by High-Intensity Laser Pulses, *Phys. Rev. Lett.* **97**, 045005 (2006).

[32] T. Esirkepov, M. Borghesi, S. V. Bulanov, G. Mourou, T. Tajima, Highly Efficient Relativistic-Ion Generation in the Laser-Piston Regime, *Phys. Rev. Lett.* **92**, 175003 (2004).

[33] A. P. L. Robinson, M. Zepf, S. Kar, R. G. Evans, C. Bellei, Radiation pressure acceleration of thin foils with circularly polarized laser pulses, *New J. Phys.* **10**, 013021 (2008).

[34] X. Q. Yan, C. Lin, Z. M. Sheng, Z. Y. Guo, B. C. Liu, Y. R. Lu, J. X. Fang, J. E. Chen, Generating High-Current Monoenergetic Proton Beams by a CircularlyPolarized Laser Pulse in the Phase-StableAcceleration Regime, *Phys. Rev. Lett.* **100**, 135003 (2008).

[35] A. Macchi, S. Veghini, F. Pegoraro, "Light Sail" Acceleration Reexamined, *Phys. Rev. Lett.* **103**, 085003 (2009).

[36] B. Qiao, M. Zepf, M. Borghesi, B. Dromey, M. Geissler, A. Karmakar, P. Gibbon, Radiation-Pressure Acceleration of Ion Beams from Nanofoil Targets: The Leaky Light-Sail Regime, *Phys. Rev. Lett.* **105**, 155002 (2010).

[37] A. Henig, S. Steinke, M. Schnürer, T. Sokollik, R. Hörlein, D. Kiefer, D. Jung, J. Schreiber, B. M. Hegelich, X. Q. Yan *et al.*, Radiation-Pressure Acceleration of Ion Beams Driven by Circularly Polarized Laser Pulses, *Phys. Rev. Lett.* **103**, 245003 (2009).

[38] C. Scullion, D. Doria, L. Romagnani, A. Sgattoni, K. Naughton, D. R. Symes, P. McKenna, A. Macchi, M. Zepf, S. Kar *et al.*, Polarization Dependence of Bulk Ion Acceleration from Ultrathin Foils Irradiated by High-Intensity Ultrashort Laser Pulses, *Phys. Rev. Lett.* **119**, 054801 (2017).

[39] F. Pegoraro, S. V. Bulanov, Photon Bubbles and Ion Acceleration in a Plasma Dominated by the Radiation Pressure of an Electromagnetic Pulse, *Phys. Rev. Lett.* **99**, 065002 (2007).

[40] Y. Wan, I. A. Andriyash, W. Lu, W. B. Mori, V. Malka, Effects of the Transverse Instability and Wave Breaking on the Laser-Driven Thin Foil Acceleration, *Phys. Rev. Lett.* **125**, 104801 (2020).

[41] F. Dollar, C. Zulick, A. G. R. Thomas, V. Chvykov, J. Davis, G. Kalinchenko, T. Matsuoka, C. McGuffey, G. M. Petrov, L. Willingale *et al.*, Finite Spot Effects on Radiation Pressure Acceleration from Intense High-Contrast Laser Interactions with Thin Targets, *Phys. Rev. Lett.* **108**, 175005 (2012).

[42] T. Toncian, M. Borghesi, J. Fuchs, E. d'Humières, P. Antici, P. Audebert, E. Brambrink, A. Cecchetti Carlo, A. Pipahl, L. Romagnani *et al.*, Ultrafast Laser-Driven Microlens to Focus and Energy-Select Mega-Electron Volt Protons, *Science* **312**, 410 (2006).

[43] K. Markey, P. McKenna, C. M. Brenner, D. C. Carroll, M. M. Günther, K. Harres, S. Kar, K. Lancaster, F. Nürnberg, M. N. Quinn *et al.*, Spectral Enhancement in the Double Pulse Regime of Laser Proton Acceleration, *Phys. Rev. Lett.* **105**, 195008 (2010).

[44] A. A. Gonoskov, A. V. Korzhimanov, V. I. Eremin, A. V. Kim, A. M. Sergeev, Multicascade Proton Acceleration by a Superintense Laser Pulse in the Regime of Relativistically Induced Slab Transparency, *Phys. Rev. Lett.* **102**, 184801 (2009).

[45] W. P. Wang, B. F. Shen, H. Zhang, X. M. Lu, J. F. Li, S. H. Zhai, S. S. Li, X. L. Wang, R. J. Xu, C. Wang *et al.*, Spectrum tailoring of low charge-to-mass ion beam by the triple-stage acceleration mechanism, *Phys. Plasmas* **26**, 043102 (2019).

[46] Y. Nodera, S. Kawata, N. Onuma, J. Limpouch, O. Klimo, T. Kikuchi, Improvement of energy-conversion efficiency from laser to proton beam in a laser-foil interaction, *Phys. Rev. E* **78**, 046401 (2008).

[47] M. Chen, A. Pukhov, T. P. Yu, Z. M. Sheng, Enhanced Collimated GeV Monoenergetic Ion Acceleration from a Shaped Foil Target Irradiated by a Circularly Polarized Laser Pulse, *Phys. Rev. Lett.* **103**, 024801 (2009).

[48] T.-P. Yu, A. Pukhov, G. Shvets, M. Chen, Stable Laser-Driven Proton Beam Acceleration from a Two-Ion-Species Ultrathin Foil, *Phys. Rev. Lett.* **105**, 065002 (2010).

[49] H. B. Zhuo, Z. L. Chen, W. Yu, Z. M. Sheng, M. Y. Yu, Z. Jin, R. Kodama, Quasimonoenergetic Proton Bunch Generation by Dual-Peaked Electrostatic-Field Acceleration in Foils Irradiated by an Intense Linearly Polarized Laser, *Phys. Rev. Lett.* **105**, 065003 (2010).

[50] T. Bartal, M. E. Foord, C. Bellei, M. H. Key, K. A. Flippo, S. A. Gaillard, D. T. Offermann, P. K. Patel, L. C. Jarrott, D. P. Higginson *et al.*, Focusing of short-pulse high-intensity laser-accelerated proton beams, *Nat. Phys.* **8**, 139 (2012).

[51] J. H. Bin, M. Yeung, Z. Gong, H. Y. Wang, C. Kreuzer, M. L. Zhou, M. J. V. Streeter, P. S. Foster, S. Cousens, B. Dromey *et al.*, Enhanced Laser-Driven Ion Acceleration by Superponderomotive Electrons Generated from Near-Critical-Density Plasma, *Phys. Rev. Lett.* **120**, 074801 (2018).

[52] D. B. Zou, A. Pukhov, L. Q. Yi, H. B. Zhuo, T. P. Yu, Y. Yin, F. Q. Shao, Laser-Driven Ion Acceleration from Plasma Micro-Channel Targets, *Sci. Rep.* **7**, 42666 (2017).

[53] F. Mackenroth, A. Gonoskov, M. Marklund, Chirped-Standing-Wave Acceleration of Ions with Intense Lasers, *Phys. Rev. Lett.* **117**, 104801 (2016).

[54] A. V. Brantov, E. A. Govras, V. F. Kovalev, V. Y. Bychenkov, Synchronized Ion Acceleration by Ultraintense Slow Light, *Phys. Rev. Lett.* **116**, 085004 (2016).

[55] P. Hilz, T. M. Ostermayr, A. Huebl, V. Bagnoud, B. Borm, M. Bussmann, M. Gallei, J. Gebhard, D. Haffa, J. Hartmann *et al.*, Isolated proton bunch acceleration by a petawatt laser pulse, *Nat. Commun.* **9**, 423 (2018).

[56] R. Matsui, Y. Fukuda, Y. Kishimoto, Quasimonoenergetic Proton Bunch Acceleration Driven by Hemispherically Converging Collisionless Shock in a Hydrogen Cluster Coupled with Relativistically Induced Transparency, *Phys. Rev. Lett.* **122**, 014804 (2019).

[57] S. Kar, H. Ahmed, R. Prasad, M. Cerchez, S. Brauckmann, B. Aurand, G. Cantono, P. Hadjisolomou, C. L. S. Lewis, A. Macchi *et al.*, Guided post-acceleration of laser-driven ions by a miniature modular structure, *Nat. Commun.* **7**, 10792 (2016).

[58] X. F. Shen, A. Pukhov, B. Qiao, Monoenergetic High-Energy Ion Source via Femtosecond Laser Interacting with a Microtape, *Phys. Rev. X* **11**, 041002 (2021).

[59] T. D. Arber, K. Bennett, C. S. Brady, A. Lawrence-Douglas, M. G. Ramsay, N. J. Sircombe, P. Gillies, R. G. Evans, H. Schmitz, A. R. Bell *et al.*, Contemporary particle-in-cell approach to laser-plasma modelling, *Plasma Phys. Controlled Fusion* **57**, 113001 (2015).

[60] K. A. Flippo, E. d'Humières, S. A. Gaillard, J. Rassuchine, D. C. Gautier, M. Schollmeier, F. Nürnberg, J. L. Kline, J. Adams, B. Albright *et al.*, Increased efficiency of short-pulse laser-generated proton beams from novel flat-top cone targets, *Phys. Plasmas* **15**, 056709 (2008).

[61] T. Matsuoka, S. Reed, C. McGuffey, S. S. Bulanov, F. Dollar, L. Willingale, V. Chvykov, G. Kalinchenko, A. Brantov, V. Y. Bychenkov *et al.*, Energetic electron and ion generation from interactions of intense laser pulses with laser machined conical targets, *Nucl. Fusion* **50**, 055006 (2010).

[62] P. Antici, S. Gaillard, L. Gremillet, M. Amin, M. Nakatsutsumi, L. Romagnani, M. Tampo, T. Toncian, R. Kodama, P. Audebert *et al.*, Optimization of flat-cone targets for enhanced laser-acceleration of protons, *Nucl. Instrum. Methods Phys. Res. A* **620**, 14 (2010).

[63] T. Kluge, S. A. Gaillard, K. A. Flippo, T. Burris-Mog, W. Enghardt, B. Gall, M. Geissel, A. Helm, S. D. Kraft, T. Lockard *et al.*, High proton energies from cone targets: electron acceleration mechanisms, *New J. Phys.* **14**, 023038 (2012).

[64] X. H. Yang, W. Yu, H. Xu, H. B. Zhuo, Y. Y. Ma, D. B. Zou, T. P. Yu, Z. Y. Ge, Y. Yin, F. Q. Shao *et al.*, Generation of high-energy-density ion bunches by ultraintense laser-cone-target interaction, *Phys. Plasmas* **21**, 063105 (2014).

[65] O. Budrigă, E. D'Humières, Modeling the ultra-high intensity laser pulse – cone target interaction for ion acceleration at CETAL facility, *Laser Part. Beams* **35**, 458 (2017).

[66] J. J. Honrubia, A. Morace, M. Murakami, On intense proton beam generation and transport in hollow cones, *Matter Radiat. Extremes* **2**, 28 (2017).

[67] J. Zhang, W. M. Wang, X. H. Yang, D. Wu, Y. Y. Ma, J. L. Jiao, Z. Zhang, F. Y. Wu, X. H. Yuan, Y. T. Li *et al.*, Double-cone ignition scheme for inertial confinement fusion, *Phil. Trans. R. Soc. A* **378**, 20200015 (2020).

[68] R. Kodama, P. A. Norreys, K. Mima, A. E. Dangor, R. G. Evans, H. Fujita, Y. Kitagawa, K. Krushelnick, T. Miyakoshi, N. Miyanaga *et al.*, Fast heating of ultrahigh-density plasma as a step towards laser fusion ignition, *Nature* **412**, 798 (2001).

[69] T. Ma, H. Sawada, P. K. Patel, C. D. Chen, L. Divol, D. P. Higginson, A. J. Kemp, M. H. Key, D. J. Larson, S. Le Pape *et al.*, Hot Electron Temperature and Coupling Efficiency Scaling with Prepulse for Cone-Guided Fast Ignition, *Phys. Rev. Lett.* **108**, 115004 (2012).

[70] X.-L. Zhu, T.-P. Yu, Z.-M. Sheng, Y. Yin, I. C. E. Turcu, A. Pukhov, Dense GeV electron–positron pairs generated by lasers in near-critical-density plasmas, *Nat. Commun.* **7**, 13686 (2016).

[71] X.-L. Zhu, Y. Yin, T.-P. Yu, F.-Q. Shao, Z.-Y. Ge, W.-Q. Wang, J.-J. Liu, Enhanced electron trapping and γ ray emission by ultra-intense laser irradiating a near-critical-density plasma filled gold cone, *New J. Phys.* **17**, 053039 (2015).

[72] X.-L. Zhu, M. Chen, T.-P. Yu, S.-M. Weng, L.-X. Hu, P. McKenna, Z.-M. Sheng, Bright attosecond γ-ray pulses from nonlinear Compton scattering with laser-illuminated compound targets, *Appl. Phys. Lett.* **112**, 174102 (2018).

[73] A. Pukhov, Z. M. Sheng, J. Meyer-ter-Vehn, Particle acceleration in relativistic laser channels, *Phys. Plasmas* **6**, 2847 (1999).

[74] Y.-Y. Ma, Z.-M. Sheng, Y.-T. Li, W.-W. Chang, X.-H. Yuan, M. Chen, H.-C. Wu, J. Zheng, J. Zhang, Dense quasi-monoenergetic attosecond electron bunches from laser interaction with wire and slice targets, *Phys. Plasmas* **13**, 110702 (2006).

[75] A. P. L. Robinson, A. V. Arefiev, D. Neely, Generating "Superponderomotive" Electrons due to a Non-Wake-Field Interaction between a Laser Pulse and a Longitudinal Electric Field, *Phys. Rev. Lett.* **111**, 065002 (2013).

[76] S. Wilks, W. Kruer, M. Tabak, A. Langdon, Absorption of ultra-intense laser pulses, *Phys. Rev. Lett.* **69**, 1383 (1992).

[77] P. Mora, Plasma Expansion into a Vacuum, *Phys. Rev. Lett.* **90**, 185002 (2003).

[78] J. E. Crow, P. L. Auer, J. E. Allen, The expansion of a plasma into a vacuum, *J. Plasma Phys.* **14**, 65 (1975).